# Molecular Model of the Contractile Ring

D. Biron,[1] E. Alvarez-Lacalle,[2] T. Tlusty,[1] and E. Moses[1]

[1]*Department of Physics of Complex Systems, Weizmann Institute of Science, Rehovot 76100, Israel*
[2]*Department d'Estructura i Constituents de la Matèria, Facultat de Física, Universitat de Barcelona,*
*Avenida Diagonal 647, E-08028 Barcelona, Spain*

We present a model for the actin contractile ring of adherent animal cells. The model suggests that the actin concentration within the ring and consequently the power that the ring exerts both increase during contraction. We demonstrate the crucial role of actin polymerization and depolymerization throughout cytokinesis, and the dominance of viscous dissipation in the dynamics. The physical origin of two phases in cytokinesis dynamics ("biphasic cytokinesis") follows from a limitation on the actin density. The model is consistent with a wide range of measurements of the midzone of dividing animal cells.

Most animal cells divide by forming a furrow at midcell, directed perpendicular to its long axis, which deepens until the two daughter cells physically separate. This process is termed "cytokinesis." Furrow kinetics measured on many different cells reveal an initial universal linear phase, in which the furrow deepens at a constant rate until the midzone reaches approximately 10% of the initial cell diameter [1]. Then the contraction enters a second, nonlinear, phase in which different mechanisms either take over or assist to complete the separation [2,3].

The hypothesis of an actin ring that actively generates contractile forces during cytokinesis was proposed by Marsland and Landau in 1954 [4] and was later directly observed. It was shown that the sliding of actin filaments ($F$-actin) driven by filaments of myosin II motors produces the forces that contract the ring [5–10]. The widely accepted model of the structure of the ring consists of a contracting network of actin filaments. This network performs a net "purse string" action as the filaments that compose it reorganize by polymerization and disassembly [1,5,8,11]. At the final stage of the contraction, the ring disintegrates [2,9]. However, a detailed elaboration of the molecular mechanism underlying the contraction has yet to be developed.

Measurements on various adherent cells show that the structure of the contractile ring is rather universal [9], and we therefore refer to a "typical" actin ring. This model ring has an initial width (the cell diameter) of $w \approx$ 10–40 $\mu$m, which contracts at a constant rate of $\dot{w} \approx$ 0.1–0.5 $\mu$m/sec (see Fig. 1). Its length is $\ell \approx$ 5–10 $\mu$m and its thickness is $\delta \approx$ 0.1–0.2 $\mu$m, both remaining approximately constant during the linear contraction phase [1,5–7,9,10,12–14]. We further assume the height of the cell does not change significantly during this time as is indicated experimentally ($h \approx$ 3–5 $\mu$m), but this assumption can be relaxed.

A typical ring contains about $10^5$ actin filaments (of diameter $a \sim$ 6–8 nm and length 0.2–0.6 $\mu$m) and about $10^4$ myosin bipolar filaments of diameter $b \sim$ 10–20 nm and length 0.5–1 $\mu$m [9,13–15]. The total volume of actin filaments is therefore about 5% of the initial ring volume, which is comparable to the volume of the ring just before disintegration.

First we consider the physical processes that oppose the contracting ring, and from these estimate the power that the actin network exerts and its scaling with time. The relevant processes are adhesion of the cell to the substrate, elastic deformation of the membrane and cytoskeleton, and the viscous damping of the flow that exits the contracting furrow.

When a dividing cell adheres tightly to a substrate, the geometry of the neck joining the two daughter cells may be approximated as an oval tube. $\ell$ is the length of the tube along the symmetry axis. The cross section of the tube has a height $h$ (perpendicular to the substrate), has a width $w$, and is stadiumlike, i.e., a straight section capped by semicircles of radius $h/2$ at its two ends (Fig. 1). The width contracts at a constant speed $\dot{w}$ while $h$ and $\ell$ remain fixed. The tube walls are lined by a contracting acto-myosin ring of thickness $\delta$.

In this geometry, the power against adhesion scales as the substrate area exposure rate $P_A \simeq \sigma \ell \dot{w}$ where $\sigma$ is the adhesion energy per unit area. The membrane bending elasticity is neglected due to its small bending modulus.

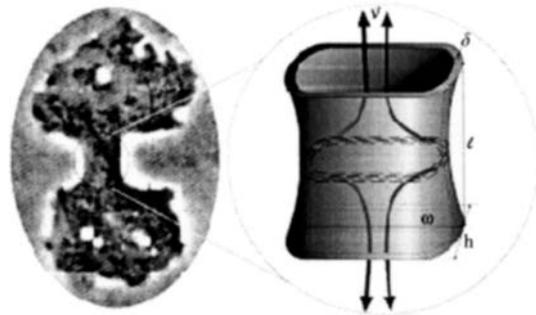

FIG. 1. Modeling the contraction of a dividing *D. discoideum* cell. Left: Microscope image of a dividing *D. Discoideum*. Right: We approximate the contractile ring by an oval tube of thickness $\delta$, length $\ell$, width $w$, and height $h$. The typical length of the cell at this stage is 20–50 $\mu$m. Dashed thick lines represent a few actin filaments at the ring center. Thick arrows depict the flow of the cytoplasm out of the tube.

The elastic deformation energy of the actin ring itself is determined by the strain in it and scales as $\mathcal{E}_E \sim E(\ell w \delta)\gamma^2$ [16] where $E$ is Young's modulus of the actin gel and $\ell w \delta$ is the volume of the ring. The strain scales as $\gamma \sim \dot{w}\tau/w$ where $\tau$ is the time it takes the actin-myosin gel to reorganize and thus release the accumulating stress [17]. The power exerted against the actin elastic forces therefore scales as $P_E \simeq \mathcal{E}_E/\tau$. The strain is limited by $\gamma \leq 1$ [17], which implies a response time $\tau \leq 10$ sec—a reasonable limit for actin contraction and reorganization.

To estimate the viscous dissipation, we need the flow velocity of the cytoplasm, which we get by comparing the outflux at the furrow ends to the change in its volume $\ell \dot{w} h \sim hwv$ (Fig. 1). The outflux velocity $v$ scales as $v \sim (\ell/w)\dot{w}$ and its gradient as $\nabla v \sim (\ell/w)(\dot{w}/h)$. The viscous dissipation is therefore $P_V = \eta \int (\nabla v)^2 dV \sim \eta \ell h w (\ell/h)^2 (\dot{w}/w)^2$ with $\eta$ the cytoplasm viscosity [18].

We evaluate the relative importance of these processes by inserting the experimentally measured values of the physical moduli, $\sigma \simeq 10$ Pa · $\mu$m [19], $E \simeq 100$ Pa [20,21], and $\eta \approx 10^4$ Pa · s [22–28]. We get $P_V/P_A \simeq (\eta \ell/\sigma)(\ell/h)(\dot{w}/w) \simeq 10^2$ and $P_A/P_E \simeq (\sigma/E\delta)\gamma^{-1} \geq 1$. We conclude that $P_V \gg P_A \geq P_E$; i.e., the relevant opposing force is due to viscosity. Moreover, the power supplied by the contracting ring scales like the viscous dissipation, which diverges as the inverse width of the furrow $P_V \sim w^{-1}$.

We simulate the contraction of the ring using the contraction of a grid in the form of a torus (see inset of Fig. 2), which adjusts dynamically as actin filaments move on it [18]. The major circumference of the torus represents the circumference of the contracting furrow of the cell. This circumference is much larger than that of the cross section of the torus, which represents the fixed thickness of the contractile ring. In practice we implement this on a doubly periodic lattice. We define large circles as trajectories in the torus that are parallel to its major circumference. They are arranged in a lattice in the cross-section plane, with either 4 or 6 nearest neighbors (for a rectangular or a hexagonal lattice, respectively). All of the large circles are assigned the same radius and the same number of mesh points.

Actin filaments move only along large circles and are $k$ mesh points long in that direction. Actin filaments are defined as neighbors if they are on neighboring circles and their centers are closer than a prescribed distance. Only neighboring filaments can interact. In reality myosin bundles are needed for this interaction, but they are not simulated explicitly. Myosin is assumed to be abundant so that it is available to mediate the interaction between any two neighboring filaments. Myosin bundles therefore affect the dynamics only through their length, which we take to be fixed and equal to $2k$. Once the centers of two actin filaments come within a distance of $l \leq 3k$ (i.e., their length plus that of the mediating myosin bundle) an attractive force begins to pull them together. This implicitly assumes that the neighboring filaments are antiparallel, an assumption that can be relaxed.

Actin filaments have two alternative modes of movement, depending on whether or not they have a neighbor. At each step, actin filaments that do not have neighbors move one mesh point either left or right with equal probability. Filaments that do have neighbors can perform a contraction step by moving one mesh point towards their neighbor. The filament makes a contraction step with probability $p$ or moves away from its neighbor with probability $q = 1 - p \leq p$. As the two filaments approach each other, $p$ increases inversely with their distance. This simulates the increase in the strength of the bond between two actual filaments as the overlap with the myosin bundle increases. Once two filaments reach $l = k$ they stop attracting each other. A sweep consists of a number of independent steps that is equal to the total number of filaments.

When actin filaments attract, they generate freed space on the lattice. Once a slice (one mesh point thick) in the cross-section plane of the torus contains no actin filaments, it can be deleted. The actual number of slice deletions is limited at every sweep by the average number of contraction steps per circle that occurred in that sweep. As a result of each deletion the major circumference of the torus is shortened by one mesh point. In reality the ring is connected to the cell membrane by integrins, and excess membrane and cytoplasm flow out of the contracting region. Deleting actin free slices from our lattice models the effects of this flow.

We use two measurements (Fig. 2) to verify that the dynamics of our model are, indeed, consistent with the scaling we derived above and with the observed behavior of contracting cells. The circumference of the ring, repre-sented by the number of mesh points on the large circles, decreases linearly in time for the majority of the process (until about 80%–90% of the contraction is completed). This is consistent with measurements of the first part of cytokinesis.

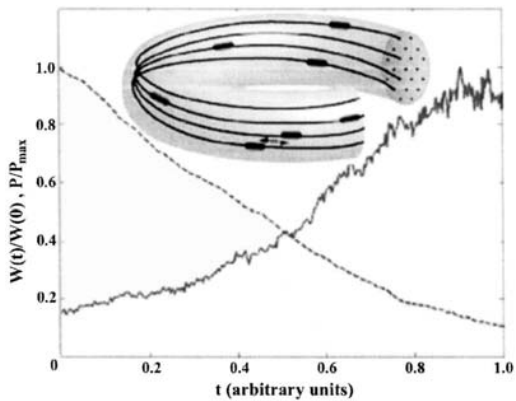

FIG. 2. Results from the simulation of the actin contractile ring. The dashed line represents the normalized furrow width, $w(t)/w(0)$. The solid line represents the normalized power $P(t)/P_{\max}$, which typically increases sixfold to tenfold. Inset: a schematic picture (not to scale) of the simulated ring. A few major circles are represented by the solid lines and their hexagonal lattice arrangement is shown on the cross section. Actin filaments, depicted as elongated ovals, can move only along the circles.

The second measurement is of the power exerted by the motors, which is calculated as the sum over all neighboring filaments of the power per contraction step. The physical origin of this power is the action of the myosin motor heads, which pull the actin filaments together. Since each myosin bundle contains many individual motor heads, the number of motors that are actually interacting with an actin filament is proportional to the overlap between the filament and the myosin bundle. The power per contraction step therefore increases in inverse proportion to the distance between the centers of the two contracting filaments.

We next construct a continuous 1d phenomenological model for the dynamics of F-actin in the ring. A previous comparable model [29] thoroughly describes the spatial organization of the ring. We expand on these ideas to treat contraction explicitly.

Let $\rho(x, t)$ be the concentration of actin filaments at time $t$ and position $x$ along a 1d axis, and let $\mathbf{C(t)} = (\pi - 2)h + 2w(t)$ be the circumference of the ring. The dimensionless normalized number of filaments, whose dynamics we analyze, is $n(x, t) = \rho(x, t)\mathbf{C(t)}$. This total number of actin filaments is conserved.

Treadmilling of randomly oriented filaments provides both fluctuations and a nonthermal diffusive mechanism [30]. The fluctuations cannot be neglected *a priori* due to the small size of the system. They are accounted for by an effective noise term, $\xi(x, t)$, whose characteristics remain to be determined. The diffusion is modeled with a diffusion coefficient, $D$.

The increase of the attraction between filaments as their overlap grows can be modeled by a subdiffusive term in a diffusionlike equation. Without fluctuations this reads

$$\partial_t n = \partial_x[D\partial_x n - P_0\Omega(\rho/\rho_0)\partial_x(n^\nu)] \qquad (1)$$

where $P_0$ measures the tendency of aggregation to amplify. This is reminiscent of a "pinching" process with $P_0$ the pinching coefficient. The exponent was chosen for simplicity to be $\nu = 1/2$, i.e., "ballistic pinching."

$\Omega(\bar\rho)$ is a cutoff function whose support is $\bar\rho_- \le \bar\rho \le \bar\rho_+$ (where $0.1 < \bar\rho_- < \bar\rho_+ < 10$ are parameters). It vanishes elsewhere, and connects smoothly between the different regions. It effectively suppresses attraction for filament concentrations that are either too low (filaments are too far apart) or too high (due to excluded volume considerations).

We identify the subdiffusive term with a flux of F-actin whose local velocity is $u(x, t) = \nu P_0\Omega(\rho/\rho_0)n^{\nu-2}\partial_x n$. Scaling $x$ with $\mathbf{C(0)}$, scaling time with $\mathbf{C^2(0)}/D$, and introducing the noise $\xi(x, t)$, we obtain

$$\partial_t n = \partial_x[\partial_x n - nu] + \xi(x, t), \qquad (2)$$

where $n(x, t = 0)$ is normalized to unity. Note that if $D = 0$ then small perturbations in the F-actin profile will quickly coarsen and the dynamics will come to a halt. For $\xi(x, t)$ we take a Gaussian delta-correlated time dependence with amplitude $I$, assuming that the dynamics of the noise has the fastest relevant time scale [18].

The contraction of the ring is obtained by integrating over all pairs of attracting filaments: $\partial_t \mathbf{C} = -\alpha \int |u\rho|dx$. $\alpha$ is a dimensionless contraction rate, and the modulus arises from the fact that motion in both the left and the right directions contributes to contraction.

We also compute the power exerted by the ring as a product of the force driving the actin filaments ($F$) and their velocity ($u$). Since the average overlap between actin and myosin is linear with the concentration of F-actin, we obtain $F \sim (\rho/\rho_0)f$, with $f(u)$ the time averaged force per myosin cross bridge. This force-velocity relation can be derived, e.g., from the power-stroke model [30]. We thus obtain $\mathbf{P} = \int (f(u)u\rho/\rho_0)\rho dx$.

We solve Eq. (2) numerically in the linearly stable configuration ($\phi = \nu P_0/D < 1$) since otherwise the system is unstable to short wavelength perturbations. Without fluctuations ($I = 0$) the dynamics enters a nonphysical steady state in which $\rho$ is constant before substantial contraction takes place, implying that the intensity of the noise must be strictly positive. In particular, when fluctuations vanish ($I \to 0$), so does contraction. The general features of the observables $\mathbf{C(t)}$ and $\mathbf{P(t)}$ depend strongly on $\bar\rho_+$ but not on $\bar\rho_-$ (which was set to 1).

Figure 3 shows the dynamics of the normalized ring circumference, $\mathbf{C(t)}/\mathbf{C(0)}$, and the normalized power it exerts, $\mathbf{P(t)}/\mathbf{P}_{\max}$. These results reproduce the basic experimental phenomena and the results obtained in our discrete simulation. The data scale with the variable $\theta \sim I\phi^\beta \alpha t$ ($\beta \approx 1.1$), so that $I$ and $D$ (in addition to $\alpha$ and $P_0$) determine the velocity of the contraction.

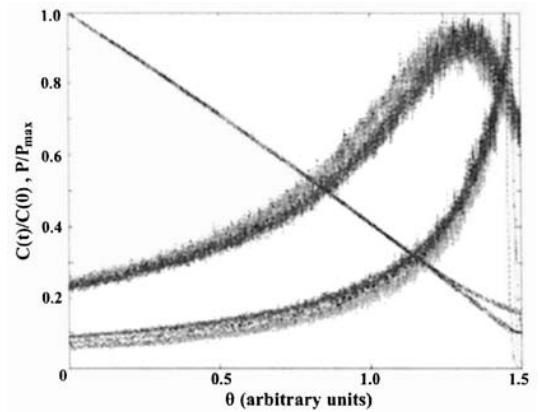

FIG. 3. Decreasing curves: $\mathbf{C(t)}/\mathbf{C(0)}$ versus $\theta \sim I\phi^\beta \alpha t$ ($\beta \approx 1.1$) for the eight combinations of $\bar\rho_+ = 2$ or 9, $I^2 = 10^{-3}$ or $5 \times 10^{-5}$, and $(\phi, \alpha) = (0.01, 1)$ or $(1, 0.01)$. The deviation from a linear behavior starts sooner for $\bar\rho_+ = 2$. Nonlinear contraction was obtained for $\phi = \alpha = 1$ (data not shown). Increasing curves: $\mathbf{P(t)}/\mathbf{P}_{\max}$ versus $\theta$ for the same parameter values. The lower (higher) family of curves is produced by $\bar\rho_+ = 9$ (2).

When $\Omega(\rho/\rho_0) \simeq 1$, then $\langle u \rangle$ is approximately constant. In this regime $C(t)$ decreases linearly and $P \sim \int \rho^2 dx \sim C^{-1} \sim w^{-1}$ increases. As $\Omega(\rho/\rho_0) \to 0$, contraction decays towards a complete halt. If the excluded volume restriction starts at low concentrations (e.g., $\bar{\rho}_+ = 2$) then $\langle u \rangle$ decreases, the contraction rate becomes sublinear relatively early [at $C(t)/C(0) \simeq 0.25$] and the power starts decreasing. If the excluded volume restriction is "rigid" (e.g., $\bar{\rho}_+ = 9$), then the linear contraction persists further until $\langle u \rangle$ and the power vanish abruptly [at $C(t)/C(0) \simeq 0.1$].

An interesting conclusion from our model is that rapid motion of filaments in random directions throughout cytokinesis is crucial for contraction. Without it, filaments will coarsen, forming separate dense bundles that are too far apart to interact and promote further contraction. It is important to note that thermal diffusion of $F$-actin is suppressed by cross-linkers (e.g., see [20]) and does not have a noticeable effect. However, the motion caused by treadmilling can produce velocities up to the order of micrometers per second [30], which is sufficient to drive the dynamics. The necessity of treadmilling for cytokinesis was demonstrated experimentally in fission yeast [31], and is consistent with several earlier experiments [32–35].

Experiments in the literature [1,12] indicate that the concentration of $F$-actin in the contractile ring indeed increases in the linear phase of cytokinesis, as we suggest. In addition, a recent atomic force microscopy measurement [36] showed that the Young modulus of the equatorial region of a dividing fibroblast monotonically increases by nearly an order of magnitude at this stage (consistent with the accumulation of $F$-actin in our model).

Several clear predictions of our model can be tested experimentally. Measuring the dynamics of actin concentration in the ring during cytokinesis should be repeated with different cells and better probes. Hopefully, one can also check whether dense actin patches form at the final stages of the linear contraction phase. The dominance of viscous dissipation implies a dependency of the dynamics on the effective viscosity of the cytoplasm. Forces (and hence power) can be measured directly during cytokinesis. The effect of additional mutations that suppress or enhance actin polymerization and depolymerization rates on the dynamics of the ring is highly interesting. Finally, measuring the degree of reorganization (e.g., by myosin motors) of actin filaments in the ring during contraction is both experimentally feasible and has implications on the power generation of this protein machinery.

This work was supported by Binational Science Foundation Grant No. 2000298 and by the EU PHYNECS network.